\documentclass[graybox, secnum]{svmult}

\usepackage{mathptmx}       
\usepackage{helvet}         
\usepackage{courier}        
\usepackage{type1cm}        
\usepackage{makeidx}         
\usepackage{graphicx}        
\usepackage{multicol}        
\usepackage[bottom]{footmisc}
\usepackage{hyperref}        
\usepackage{soul}            
\hypersetup{colorlinks=true,urlcolor=blue}
\usepackage[square,numbers]{natbib}
\makeindex             
\usepackage{amsmath}

\def\vp{\varphi_{+}}
\def\vm{\varphi_{-}}
\def\vpm{\varphi_{\pm}}

\def\dvp{{\dot \varphi}_{+}}
\def\dvm{{\dot \varphi}_{-}}
\def\dvpm{{\dot \varphi}_{\pm}}

\def\bvp{{\bar \varphi}_{+}}

\def\d{{\rm d}}
\def\dd{{\rm d}}

\def\beq{\begin{equation}}
\def\eeq{\end{equation}}
\def\bea{\begin{eqnarray}}
\def\eea{\end{eqnarray}}
\def\E{{\cal E}}

\def\d{{\rm d}}
\def\dd{{\rm d}}

\def\O{{\cal O}}

\def\Mpl{M_{\rm pl}}

\def\d{{\rm d}}

\def\I{{\cal I}}
\def \K {{k}}

\def\k{{\vec{k}}}
\def\ksub{{\vec{k}}}

\def\x{{\vec x}}

\def\t{\texttt{t}}
\newcommand{\s}{\hspace{0.8pt}}

\def\beq{\begin{equation}}
\def\eeq{\end{equation}}

\def\bea{\begin{eqnarray}}
\def\eea{\end{eqnarray}}

\begin{document}
\title*{EFT for de Sitter Space}
\author{Daniel Green}
\institute{Daniel Green \at UC San Diego, La Jolla, USA, \email{drgreen@physics.ucsd.edu}}
%
%
\maketitle
\abstract{The physics of de Sitter space is essential to our understanding of our cosmological past, present, and future.  It forms the foundation for the statistical predictions of inflation in terms of quantum vacuum fluctuations that are being tested with cosmic surveys.  In addition,  the current expansion of the universe is dominated by an apparently constant vacuum energy and we again find our universe described by a de Sitter epoch. Despite the success of our predictions for cosmological observables, conceptual questions of the nature of de Sitter abound and are exacerbated by technical challenges in quantum field theory and perturbative quantum gravity in curved backgrounds. In recent years, significant process has been made using effective field theory techniques to tame these breakdowns of perturbation theory. We will discuss how to understand the long-wavelength fluctuations produced by accelerating cosmological backgrounds and how to resolve both the UV and IR obstacles that arise.  Divergences at long wavelengths are resummed by renormalization group (RG) flow in the EFT. For light scalar fields, the RG flow manifests itself as the stochastic inflation formalism. In single-field inflation, long-wavelength metric fluctuations are conserved outside the horizon to all-loop order, which can be understood easily in EFT terms from power counting and symmetries.}

\section*{Keywords} 
Effective Field Theory, de Sitter space, Inflation, Cosmology

\section{Introduction}

The physics of de Sitter space forms an important pillar in our understanding of the universe.  Structure in the universe is widely believed to have originated in the distance past during an approximately de Sitter phase, inflation, where the tools of quantum field theory (QFT) in curved space are essential for computing the statistical predictions of inflation. While tree-level calculations are in precise agreement with observational data, the theoretical foundations of these calculations are poorly developed compared to their analogues in flat space. These complications persist in our in attempts to understand the universe at late-times. Observations of the expansion of the universe at low redshifts are consistent with the existence of a new phase de Sitter-like expansion. Our understanding of the universe in the current epoch is limited by the same technical complications as inflation and by a number of open questions about the origin and fate of the small non-zero vacuum energy.

In classical general relativity, de Sitter (dS) space presents few mysteries~\cite{Spradlin:2001pw}.  It is a maximally symmetric solution to Einstein's equations, exhibiting a $SO(d,1)$ group of isometries in $d$ spacetime dimensions.  For understanding out own universe, we often focus on the patch of de Sitter described in terms of an expanding FRW solution, 
\beq
ds^2 = g_{\mu \nu}^{\rm dS} dx^\mu dx^\nu = -dt^2 + a(t)^2 d\x^2 = a(\tau)^2 (- d\tau^2 + d\x^2) \ , 
\eeq
where $a(t) =e^{Ht}$ or $a(\tau) H = -1/\tau$.  

The quantum nature of de Sitter space is significantly more complicated, even for non-interacting quantum fields. The litany of conceptual and technical challenges begin with the ultraviolet (UV) limit, where the so-called {\it trans-Planckian} problem~\cite{Brandenberger:1999sw,Starobinsky:2001kn,Brandenberger:2004kx} suggests that even our choice of vacuum might be sensitive to Planck-scale physics. The questions persist as we move to long distances, where infrared (IR) divergences and secular growth challenge our notion of a perturbative expansion~\cite{Ford:1984hs,Antoniadis:1985pj,Tsamis:1994ca,Tsamis:1996qm,Seery:2010kh,Giddings:2010nc,Burgess:2010dd,Rajaraman:2010xd,Marolf:2010zp,Hu:2018nxy,Akhmedov:2019cfd}. This problem is compounded at higher-loop order where our treatment of the UV and IR regimes manifests itself as additional divergences in loop diagrams.   The predictions for the scalar metric fluctuations that sourced structure in the universe are sensitive to how we treat these regimes; therefore, our belief that inflation is consistent with observations is dependent on a satisfactory resolution to these challenges.

Fortunately, effective field theory (EFT) ~\cite{Polchinski:1992ed,Georgi:1993mps,Manohar:2018aog,Cohen:2019wxr} provides a variety of tools to understand both the short and long distance behavior of de Sitter space. Fundamentally, the difficulty with cosmological backgrounds is that UV physics evolves to the IR through the expansion of the universe.  This process does not violate the fundamental principles of EFT, like the decoupling of scales, but it does make their manifestation less transparent. Of course, we should expect these principles to hold in dS, given that our universe is currently in a dS-like phase and we use EFT successfully to describe the world around us.  Yet, it remains challenging to make sense of decoupling in contexts, like cosmological particle production, where the curvature of spacetime is essential to the physical process.  Our goal in this chapter is to demonstrate that QFT and perturbative quantum gravity in dS can be recast in the familiar language of EFT where the resolutions to many of these problems have ready-made solutions.

The most basic insight that underlies the success of EFT is that the physics de Sitter is characterized by a single energy scale, $H$, the Hubble scale.  First and foremost, the blue-shifting of modes as we evolve backwards in time does not negatively impact perturbation theory (in the Bunch-Davies vacuum, or a finite energy excitation thereof) because physical processes do not occur at these high energies.  Instead, the energy scale associated with particle production is $H$, such that Planckian physics is exponentially suppressed when $H \ll \Mpl$, where $\Mpl$ is the reduced Planck mass. This observation is well-known, especially in inflationary phenomenology, where the amplitude of primordial non-Gaussianity is usually determined by power counting in $H$~\cite{Cheung:2007st}.  The implications are less obvious in loop diagrams, where one encounters logarthmic divergences that need to be regulated. However, if one defines the strength of couplings at the Hubble scale, the familiar RG from flat space is unnecessary and all such logarithmic terms vanish. Many of these observations may even seem self-evident but can become obscured without effective regulators in dS~\cite{Senatore:2009cf}. 

The second, and less obvious, outcome of the EFT approach to de Sitter~\cite{Cohen:2020php}, is that the super-horizon evolution of fields and composite operators can be organized by explicit power counting in powers of $k /(aH) \ll 1$. In the process, the degrees of freedom are redefined according to their scaling dimension in $k$, and time evolution is recast in the language of dynamical RG flow, so that the powers of $k/(aH)$ that appear follow from dimensional analysis. The EFT does not contain any relevant operators (in the RG sense), reproducing the long known result that corrections grow at most logarithmically~\cite{Weinberg:2005vy,Weinberg:2006ac}. Logarithmic terms can be understood as operator mixing, while irrelevant terms decouple at late times. For massless scalars, an infinite number of operators can mix, giving rise to the framework of stochastic inflation as the master equation for this RG flow.  For metric fluctuations, the all-orders conservation of the adiabatic and tensor metric fluctuations follow from power counting, as the dimensions of these operators are fixed by symmetries and cannot be modified by RG.


The results in this chapter will be presented from the point of view of EFT, particularly Soft de Sitter Effective Theory (SdSET)~\cite{Cohen:2020php}, applied to (in-in) cosmological correlators of scalar fields in fixed dS and metric fluctuations in single-field inflation. Many of the key results have been or can be derived from different perspectives, including conventional perturbation theory~\cite{Baumgart:2019clc,Baumgart:2020oby}, the wavefunction of the universe~\cite{Gorbenko:2019rza}, and/or the physics of the static patch~\cite{Mirbabayi:2019qtx,Mirbabayi:2020vyt}. SdSET has the unique advantage that many non-trivial results when explained in terms of diagrams of the original theory, become simple observations about dimensional analysis within the EFT. In addition, hard to interpret IR divergences in the full theory are traded for UV divergences in the EFT where they have a standard interpretation in terms of RG. Our emphasis on SdSET is similar to the role of the exact RG and EFT in Polchinski's proof of renormalizability of $\lambda \phi^4$ in flat space~\cite{Polchinski:1983gv}; although one can reach the main result by diagrammatic arguments~\cite{Weinberg:1959nj}, Polchinski's exact RG makes the conceptual meaning of the result transparent, and generalizes it to other theories. Our point of view in this chapter, as with many presentations of EFT, is that we will only claim to have fully understood a phenomona when it can be explained by symmetries and power counting. 


This chapter will be organized as follows: In Section~\ref{sec:EFT}, we will discuss dS as an EFT where the relevant energy scale is the expansion rate $H \ll \Mpl$.  Perturbation theory will be controlled by the small size of the expansion rate in a predictable way.  We will specifically show how there is no trans-Planckian problem~\cite{Brandenberger:1999sw,Starobinsky:2001kn,Brandenberger:2004kx} unless we give up the idea that short distance physics of de Sitter is similar to flat space (which would also contradict everyday experience).  In Section~\ref{sec:sdset}, we discuss how to understand inflationary and dS backgrounds on scales much larger than the size of the cosmological horizon, $H^{-1}$. These are the scales that give rise to IR divergences and secular growth in traditional approaches to perturbation theory.  We will introduce the SdSET to handle this regime and we will see that the IR divergences of the full theory are replaced with EFT UV divergences, so that they can be resummed via renormalization group (RG) flow, following the usual EFT playbook. In Section~\ref{sec:light}, we apply these results to massless scalars and show how stochastic inflation arises from operator mixing in SdSET.  In Section~\ref{sec:metric}, we then explain how the all-orders conservation of the metric follows from power counting and discuss some implications for slow-roll eternal inflation. We conclude in Section~\ref{sec:conclusions}.

\section{Effective Theory in de Sitter}\label{sec:EFT}

All discussions of de Sitter space start from the central premise that the curvature of dS is small in Planck units or $H \ll \Mpl$.  This ensures that our classical solution for the background is under control and can be described geometrically, up to small perturbations. Without such an assumption, there is no controlled background geometry in which to discuss quantum fields or metric fluctuations. However, even with this assumption, it is still not necessarily obvious that quantum gravitational effects are always suppressed by at least $(H/\Mpl)^2$. 

The most unambiguous way to define the energies relevant to a given process is to calculate an observable quantity sensitive to the physical energy scale.  To simplify the discussion, we will mostly consider the case of a scalar field $\phi$ of mass $m$ and action (in the FRW slicing) \ ,
\beq\label{eq:massive_scalar}
S = \int dt d^3x \, a^3(t) \left[ \frac{1}{2} \dot \phi^2 - \frac{1}{a(t)^2} \partial_i\phi \partial^i \phi -  m^2 \phi^2 \right] \ ,
\eeq
where the spatial indices are raised with the Kronecker $\delta^{ij}$. Following standard canonical quantization, we decompose the field in terms of fourier modes, $\k$, according to
\beq\
\phi(\x,\tau) = \int \frac{\dd^3 k}{(2\s\pi)^3}\s e^{i\s \ksub\cdot \x} \left(\bar \phi^*\big(\s \k,\tau\big) a_{\k}^\dagger + \bar \phi\big(\s \k,\tau\big) a_{-\k} \right) \ ,
\eeq
where 
\beq\label{eq:mode}
\bar \phi\big(\s\k,\tau\big) =-i\s e^{i\s\left(\nu+\frac{1}{2}\right) \frac{\pi}{2}} \frac{\sqrt{\pi}}{2} H(-\tau)^{3 / 2} H_{\nu}^{(1)}(-k\s \tau) \ ,
\eeq
is a solution to the classical equations of motion, where $\nu = \sqrt{\frac{9}{4}-\frac{m^2}{H^2}}$ and $H_\nu^{(1)}$ is the Hankel function of the first kind. Next, we promote $a_{\k}^\dagger$ and $a_\k$ to quantum mechanical operators that act on the vacuum state.  The choice of vacuum is often presented as an ambiguity unique to de Sitter; however, to be consistent with physical expectations and experience in our own universe, we will require that when wavelength of the modes is subhorizon, $k \gg aH$, we reproduce the vacuum of flat space.  Specifically, this means that as $\tau \to -\infty$, $\phi$ should behave as a field operator in flat space, namely that $a_{\k}^\dagger$ creates a particle from the vacuum while $a_\k$ annihilates the vacuum. We can see that Equation~(\ref{eq:mode}) is a negative frequency mode, as needed, by expanding the Hankel function in $\tau \to -\infty$ we find
\beq
\bar \phi \to -i \frac{H(-\tau)}{\sqrt{2 k}} \exp\left( -i k \tau \right)  \ .
\eeq 
This takes the form of a WKB solution for negative frequency mode if we identify the physical (WKB) frequency as
\beq
\omega_{\rm physical}(t)= \frac{k}{a(t)} \quad \to \quad \int^tdt' \omega(t') =  \int^t dt' \frac{k}{a(t')} = - \frac{k}{a(t)H} = k\tau  \ .
\eeq
In this sense, using the canonical commutation relation for $a_\k$ and $a^\dagger_\k$, 
\beq
\big[a_\ksub^{\dagger}\s,\s a_{\k'} \big] = (2\s \pi)^3\s \delta\big(\k-\k' \big)  \qquad a_\k |0 \rangle =  \langle 0 | a^\dagger_\k  = 0 \ ,
\eeq
reproduced flat space physics on short distances where $k \gg aH$.  We may choose other states, corresponding to excitations of the flat space vacuum. For most such choices, the energy density also breaks the de Sitter symmetry and redshifts away through the expansion of the universe.  The exceptions are the $\alpha$-vacua~\cite{Allen:1985ux,Mottola:1984ar}, which are de Sitter invariant but correspond to infinite energy configurations from the flat space perspective and may be ill-defined when including interactions~\cite{Banks:2002nv}. Arguably the most important non-trivial excited state for the purpose of cosmology is a time-dependent background for the scalar field, $\phi(\x,t) = \phi_0(t)$.  A case of particular interest is inflation~\cite{Baumann:2009ds}, where the energy density of the time evolution is well above the Hubble scale, $\dot \phi_0^2 \gg H^4$.  

For cosmological applications (see e.g.~\cite{Baumann:2009ds}), we are particularly interested in the case of massless scalars, $m^2=0$ ($\nu = 3/2$) where 
\beq
\bar \phi\big(\s\k,\tau\big) \to \frac{H}{\sqrt{2 k^3}} (1-i k \tau) e^{i k \tau} \ ,
\eeq
and the equal-time two-point function for super-horizon modes ($k \tau \ll 1$) becomes
\beq\label{eq:power}
\langle \phi(\k, \tau)\phi(\k', \tau) \rangle = \frac{H^2}{2 k^3 } (2\pi)^3 \delta(\k + \k') \ .
\eeq
Although this is a quantum-mechanical calculation, the commutator $[\dot \phi, \phi]\propto a^{-3}(t)$ vanishes at long wavelengths and $\phi(-k \tau \ll 1)$ becomes an effectively classical statistical fluctuation. Essentially the same behavior is found for the tensor fluctuations of the metric, $g_{\mu \nu} = g_{\mu \nu}^{\rm dS} +\gamma_{\mu \nu}$, and repeating this calculation gives their power spectrum
\beq
\langle \gamma^s(\k, \tau)\gamma^{s'} (\k', \tau) \rangle = \frac{2 H^2}{\Mpl^2}\frac{1}{k^3 } (2\pi)^3 \delta(\k + \k')  \delta_{s,s'} \ ,
\eeq
where $s, s'$ are the two helicities of the graviton. The small amplitude of the metric fluctuations confirms our intuition that when $H \ll \Mpl$ the metric is well described by the classical geometry, $g_{\mu \nu}^{\rm dS}$, up to small perturbations. 

Although these calculations may be consistent with our intuition, the assumptions about early times may not seem so innocuous, as often articulated in terms of the trans-Planckian problem~\cite{Brandenberger:1999sw,Starobinsky:2001kn,Brandenberger:2004kx}. The source of concern is that we are defining the vacuum of the field in the far past, $\tau \to -\infty$, where the energy of a mode diverges, $\omega = -k\tau \to \infty$.  Clearly, for any $k > 0$, there exists a time in the past where $\omega > \Mpl$ and, naively, there is a breakdown in our EFT.  While potentially concerning, this is not necessarily a problem for the following reason: the energy of a single particle is not a Lorentz invariant quantity. It is therefore not a given that when $\omega \gg \Mpl$ there must be a breakdown of EFT, as it depends on the coordinate system.  We can always go to some boosted coordinate system where then energy is below the Planck scale.  This is familiar from flat space, where EFT breaks down when a Lorentz invariant quantity, like the center of mass energy $s = (\omega_1+\omega_2)^2 - (\k_1 + \k_2)^2$, is larger than the cutoff of the EFT (the Planck scale in this case).  

The challenge in dS is that there is no obvious analogue of the Mandelstam variables $s$, $t$, $u$ that we can use to diagnose the breakdown of EFT.  Instead we can simply look at what scales appear in our cosmological observables and whether there is a well-defined expansion in a small parameter.  In the cosmological setting, we will consider\footnote{An alternate approach to cosmological correlators is to first calculate wavefunction of the universe~\cite{Hartle:1983ai} and the subsequently determine the in-in correlators by the usual rules of quantum mechanics.  This point of view has some advantages~\cite{Benincasa:2022gtd,Benincasa:2022omn} including making the connection to holography in AdS more transparent~\cite{Maldacena:2002vr,Harlow:2011ke,Maldacena:2011nz} and clarifying the origin of some IR divergences~\cite{Anninos:2014lwa,Gorbenko:2019rza}.} equal time in-in correlation functions~\cite{Weinberg:2005vy,Weinberg:2006ac}, 
\bea
&& \langle {\rm in}|  Q(t) |{\rm in} \rangle=  \nonumber \\
&& \left\langle\bar{T}\exp \left[i \int_{-\infty(1+i \epsilon)}^{t} H_{\mathrm{int}}(t') d t' \right]  \, Q_{\mathrm{int}}(t)  \, T \exp \left[-i \int_{-\infty(1-i \epsilon)}^{t} H_{\mathrm{int}}(t') d t' \right]\right\rangle \ , \label{eq:inin}
\eea 
where $H_{\rm int}(t)= - \int d^3 x a^3(t) {\cal L}_{\rm int}(\phi_{\rm int}(\x,t))$is the interaction Hamiltonian, $\phi_{\rm int}(\x,t)$ are the interaction pictures fields, and $Q(t)$ ($Q_{\rm int}(t)$) is some operator composed of $\phi(\k_i,t)$ ($\phi_{\rm int}(\k_i,t)$) with $k_i \ll aH(t)$.  At first sight, the integrals run to $t =-\infty$ and would indicate that the super-Planckian modes contribute to this correlator.  This is a red-hearing, as the integral is oscillatory and therefore these contributions can (and will) cancel out.  To see this~\cite{Behbahani:2012be}, we note that the integrals are defined by the $i \epsilon$ prescription to ensure we are in the lowest energy state. As a result, the original contour (in time) is composed of two pieces, $\tau = \tau_0 + (1\pm i \epsilon) \tau$ where $\tau \in (\pm \infty,0]$ and $\tau_0 < 0$. We can Wick rotate this contour, $(1\pm i \epsilon) \tau \to \pm i \tau_E$, to give a single anti-time-ordered correlator,
\beq
 \langle {\rm in}|  Q(t) |{\rm in} \rangle =\left\langle\bar{T}\left(  Q_{\mathrm{int}}(\tau_{0}) \exp \left[-\int_{-\infty}^{\infty} H_{\text {int }}\left(i \tau_{E}+\tau_{0}\right) a\left(i \tau_{E}+\tau_{0}\right) d \tau_{E}\right]\right)\right\rangle \ .
\eeq
We can then calculate any correlator using the anti-time-ordered Green function, $\langle \bar T \phi(\k, \tau_E) \phi(\k', \tau_E') \rangle \propto e^{- k |\tau_E-\tau_E'|}$. Taking $\tau_E \to \pm \infty$ for any one mode will always produce an exponentially suppressed contribution to the correlator and thus does not suggest a breakdown in perturbation theory. Instead, the correlators are dominated by $k|\tau_E | = {\cal O}(1)$ or energies $\omega_{\rm physical}  = {\cal O}(H)$.

\begin{figure}[t]
	\begin{center}
		\includegraphics[width=3.5in]{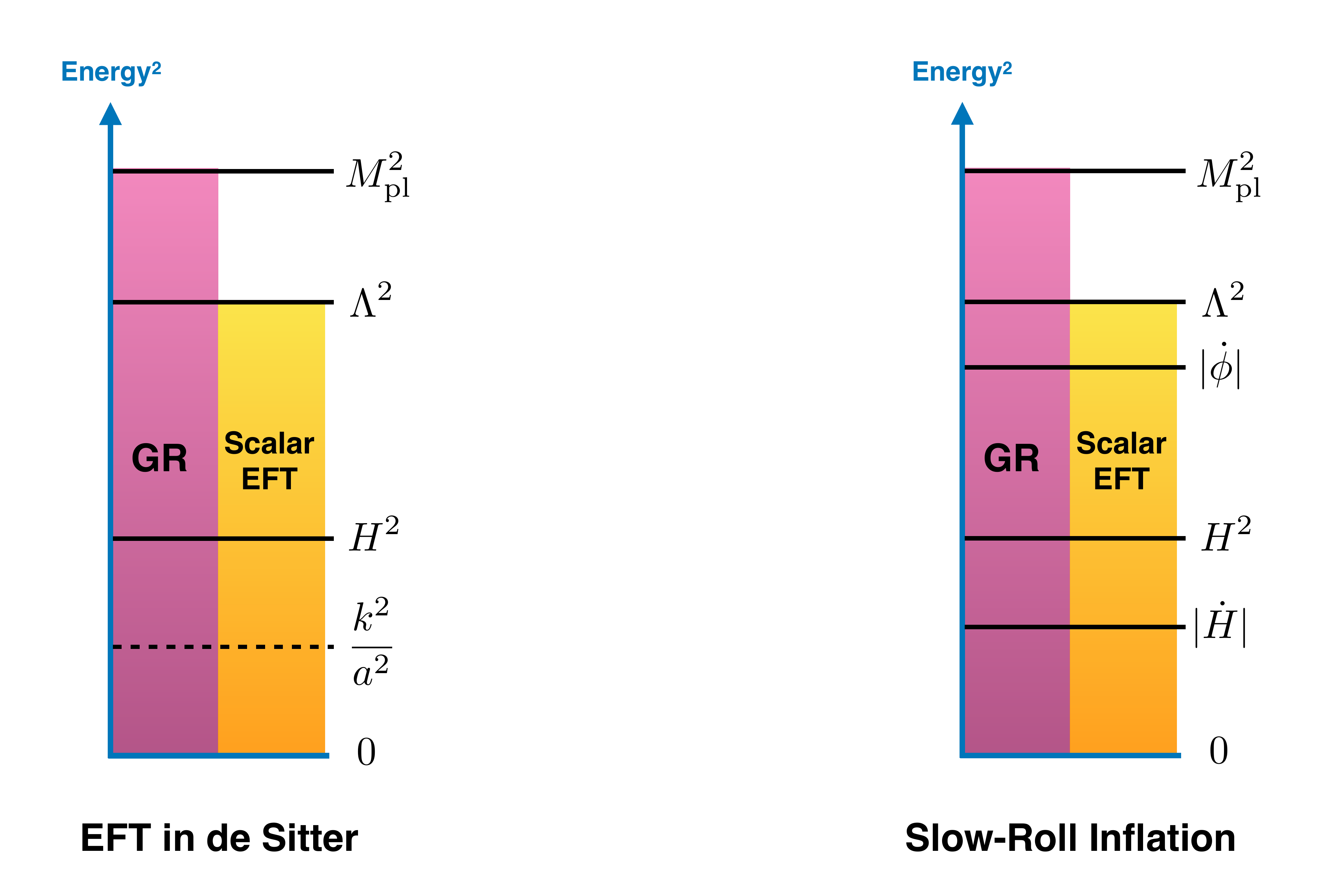}
	\end{center}
	\vskip -.5cm
	\caption{The hierarchy of scales that defines EFT in de Sitter ({\it Left}) and slow-roll inflation ({\it Right}). For the background to be under control, we always require $H^2 \ll \Mpl^2$. In pure dS, scalar EFT is under control as long as $H^2 \ll \Lambda^2$.  In contrast, in slow-roll inflation the scalar background is controlled by $|\dot \phi | \gg H^2$ and therefore requires the stronger constraint that $|\dot \phi | \ll \Lambda^2$.  This gives rise to the result that equilateral non-Gassuianity is small in slow-roll, $f_{\rm NL}^{\rm eq} \ll 1$~\cite{Creminelli:2003iq}.}
	\label{fig:EFT_scales}
\end{figure}

The intuition that the relevant energy scale is $H$ is also confirmed by explicit calculations.  For example, a massless scalar with a derivative interaction, 
\beq\label{eq:derivative_int}
{\cal L}_{\rm int} = \frac{1}{\Lambda^4}   (\partial_\mu \phi \partial^\mu \phi)^2 \ ,
\eeq
produces a four-point function
\beq\label{eq:4point}
\langle \phi(\k_1) \phi(\k_2) \phi(\k_3) \phi(\k_4) \rangle = \frac{H^8}{\Lambda^4} \frac{P_6(\k_1,\k_2,\k_3\,\k_4)}{ 16 (k_1 k_2 k_3 k_4)^3 k_t^5} (2\pi)^3 \delta(\sum \k_i) \ , 
\eeq
where $k_t= (k_1+k_2+k_3+k_4)$ and $P_6$ is a sixth order polynomial given in~\cite{Creminelli:2011mw}.  It is noteworthy that the cutoff appears in the dimensionless ratio $(H/\Lambda)^4$ which controls the amplitude of the correlator.  As a result, the correlators are approximately Gaussian when $H \ll \Lambda$ and consistent with weak coupling.  The additional powers of $H$ also arise from dimensional analysis as $\phi$ is a field of dimension one, while $k\tau$ is dimensionless.  From these two observations, it is easy to see that $H$ is the only energy scale beyond $\Lambda$ needed to make the correct units in Equation~(\ref{eq:mode}). This situation becomes more complicated in inflationary backgrounds, where $\dot \phi^2 \gg H^4$ and we therefore require $\dot \phi^2 \ll \Lambda^4$ for control at weak coupling~\cite{Creminelli:2003iq} (see Figure~\ref{fig:EFT_scales} for illustration). 

Naturally,  it seems reasonable to assume that the tools of EFT are applicable and under control in dS if the cutoff of the EFT is above the Hubble scale, $\Lambda \gg H$.  Concretely, suppose we are given an EFT with an interaction Hamiltonian density
\beq
 {\cal H}_{\rm int} =\sum_\Delta \frac{\lambda_\Delta}{ \Lambda^{\Delta-4}} {\cal O}_\Delta \ ,
\eeq
where we are working in $d=4$ spacetime dimensions, and the sum is over the list of operators, ${\cal O}_\Delta$, organized by their (flat space) scaling dimension $\Delta$. Then, at tree level, each term in the sum will contribute to a correlator whose dimensionless amplitude is $\lambda_\Delta (H/\Lambda)^{\Delta -4}$.  

The intuition that $H$ sets the physical scale of de Sitter correlators is further confirmed at one-loop. As with tree-level correlators, all the explicit units of energy are replaced with the appropriate powers of $H$, such that any large loop correlations arise from divergences in dimensionless integrals involving $k$ and/or $\tau$.  Generic loop corrections will be discussed in Section~\ref{sec:sdset}, but we can gain significant intuition from specialized to the case $\nu = 1/2$ ($m^2 = 2 H^2$), which is the conformal mass~\cite{Green:2020txs}. De Sitter space is conformally flat and for this special value of the mass, certain field theories in de Sitter are simply the Weyl transform of a conformal field theory (CFT) involving a massless\footnote{The nonzero mass in dS arises from the curvature coupling needed to make the gravitational $T^\mu_\mu =0$ for a free scalar in flat space.} scalar in flat space. Concretely, for an operator of dimension $\Delta$, the relation between the dS and CFT correlators is given by
\beq\label{eq:p2_CFT}
\langle \O_\Delta(\x, \tau) \ldots \rangle_{\rm dS} = (a(\tau))^{-\Delta} \langle \O_\Delta(\x, \tau) \ldots \rangle_{\rm CFT}  \ .
\eeq
One can calculate this explicitly in the case of $\lambda \phi^4$ in $d=4-\epsilon$ dimensions, which is conformal for $\lambda =0$ and $\lambda =\lambda_\star = \frac{16 \pi^2}{3} \epsilon$ (see e.g.~\cite{Rychkov:2015naa}).  The dimension of $\phi^2$ in flat space is $\Delta = d-2$ at $\lambda =0$, and $\Delta = d-2 +\epsilon/3$ at the nontrivial fixed point $\lambda =  \lambda_\star$. 

Using the Weyl transformation from flat space, the de Sitter correlator at the non-trivial fixed point is necessarily
\beq
\langle \phi^2(\k, \tau) \phi^2(\k', 0) \rangle_{\rm dS}  \propto \frac{k^{1+\epsilon/3}}{a(\tau)^{4-2\epsilon/3} } \ ,
\eeq
where we have defined
\bea
\phi^n(\k,\tau) &=& \int d^{3-\epsilon} x e^{-i \k \cdot \x} \phi^n(\x,\tau)  \nonumber \\
&=& \int \left (\prod_{i=1}^{n} \frac{d^{3-\epsilon} k_i}{(2\pi)^{3-\epsilon}} \phi(\k_i,\tau) \right) (2\pi)^{3-\epsilon} \delta \left( \sum_{i=1}^n \k_i - \k\right) \ .
\eea
Clearly we must get the same answer by perturbation theory in dS. By direct calculation, the leading perturbative correction in $\lambda$ gives
   \bea \label{eq:phi2phi2fullcorrelator}
     \langle \phi^2(\k,\tau)\phi^2(-\k,\tau)\rangle
     && = \frac{c}{2  a(\tau)^{2\Delta_{\phi^2}}} \ \K^{1-\epsilon}\\
&\times&      \left[
        1+\frac{\lambda}{32 \pi^4 C} \,
        \left(
        \frac{1}{\epsilon} + \log \left(\frac{\mu}{H} \right) - \log ( -\K \tau ) + \dots
        \right) \
      \right]. \nonumber
    \eea
where $\mu$ is the renormalization scale, $\Delta_{\phi^2} = d-2$, and $C \approx -1/(4\pi^2)$ is a constant. The key thing to notice is that the logarithmic divergence that can be separated into $\log \mu / H$ and $\log(- k\tau)$.  Since the metric is unchanged by the rescaling $a \to \xi a$ and $x \to x / \xi$, it is required that powers of $k$ and $a$ must appear together as the physical wavenumber $k/a$~\cite{Senatore:2009cf}. Using $\tau = -1/(aH)$ one can see that any log should be rewritten in terms of these two dimensionless ratios.  This is true more generally than this one example, of course, and can be seen a variety of loop corrections calculated in the literature~\cite{Burgess:2010dd,Anninos:2014lwa}.  It follows that any $\log \mu /H$ can be set to zero by simply choosing $\mu =H$, which means that $H$ is the correct physical scale we should associate with de Sitter correlators.  With this choice, there is no conventional renormalization group flow; there are no large logs involving the renormalization scale.  

One may be troubled that setting $\mu =H$ has left us with a large log, $\log(-k \tau)$. These are large infrared logs which we will handle by matching onto an EFT for the long modes, with $k/a \ll H$.  Yet, for the case of $\lambda \phi^4$ in $d=4-\epsilon$, we know that this log must be resummed in order to match the result from the Weyl transformation in Equation~(\ref{eq:p2_CFT}).  The procedure for doing this known as {\it dynamical RG} and involves introducing an unphysical time, $\tau_\star$, to regulate the large logs.  As with conventional RG, we define a renormalized operator ${\cal O}_{\rm R} =Z(k_\star \tau_\star) {\cal O}$ so that
   \beq \label{eq:phi2phi2_R}
     \langle \phi_{\rm R}^2(\k,\tau)\phi_{\rm R}^2(-\k,\tau)\rangle
= \frac{c}{2  a(\tau)^{2\Delta_{\phi^2}}} \ \K^{1-\epsilon}   \left[
        1+\frac{\lambda}{32 \pi^4 C} \,
      \log ( k_\star \tau_\star/(k \tau))  +\ldots \
      \right].
    \eeq
Since ${\cal O}$ is independent of $\tau_\star$, one can write a differential equation of
\beq
\frac{d}{d \log( k_\star \tau_\star)}  \langle \phi_{\rm R}^2(\k,\tau)\phi_{\rm R}^2(-\k,\tau)\rangle =  \frac{\lambda}{8 \pi^2}  \langle \phi_{\rm R}^2(\k,\tau)\phi_{\rm R}^2(-\k,\tau)\rangle 
\eeq
Once we reconganize that $k_\star \tau_\star / (k \tau)$ always appears together, we can write this as a differential equation in $k \tau$, whose solution is 
\beq\label{eq:resummed}
\langle \phi_{\rm R}^2(\k,\tau)\phi_{\rm R}^2(-\k,\tau)\rangle = \frac{1}{a(\tau)^{2\Delta_{\phi^2}+2\gamma_{\phi^2}}} \ \K^{1-\epsilon+ 2\gamma_{\phi^2}}
\eeq
where $\gamma_{\phi^2} =  \lambda_H/ 16\pi^2$, so that at the conformal fixed point $\lambda_\star =16\pi^2 \epsilon/3$, we find exact agreement between Equations~(\ref{eq:resummed}) and~(\ref{eq:p2_CFT}). Note that this procedure is nearly identical to the perturbative derivation of (\ref{eq:p2_CFT}) in flat space, where one would use the Callan-Symanzik equations to convert a logarithmic correction (the anomalous dimension) into a power-law.

These results suggest that we should think about resumming the large time-dependent logs in general~\cite{Burgess:2009bs,Seery:2010kh,Youssef:2013by,Gautier:2013aoa,Kitamoto:2018dek}, and not just at the conformal fixed point. There is nothing about the calculation that suggests it is important that we are at the fixed point, particularly as the RG itself is unimportant after we set $\mu =H$. The result of this resummation is that we introduce an anomalous dimension for $\phi^2$, $\gamma_{\phi^2} = \lambda_H/ 16\pi^2$ where the coupling $\lambda_H$ is the value of the coupling at $\mu =H$. This suggests a deeper connection between the large logs and anomalous dimensions, as we will see in the next Section. These results are also closely related to results in the cosmological bootstrap program~\cite{Baumann:2022jpr}, particularly the non-perturbative structure of QFT correlators in dS~\cite{Hogervorst:2021uvp,DiPietro:2021sjt}. We will not explore that connection here as our emphasis is EFT, particularly with an eye towards perturbative quantum gravity.

The idea of dynamical RG as a solution to the IR behavior in dS has a long history~\cite{Podolsky:2008qq,Burgess:2009bs,Dias:2012qy} and, in many cases, it was known to be the most physically sensible solution to the problem of secular growth. Yet, we should note that the not every appearance of $\log (-k \tau)$ signals some universal correction to the long distance physics; logs may also arise from finite scheme-dependent corrections to the parameters of the Lagrangian. For example, due to the challenge of finding good mass independent regulators, loop diagrams may introduce a finite shift of the mass of our field $\phi$ by $m^2  \to m^2 + \kappa H^2$ for some $\kappa ={\cal O}(\lambda)$.  As $m^2$ changes the super-horizon scaling with $a(t)$, a perturbative correction to $m^2$ will take the form of logarithmic growth.  However, unlike our usual expectations around logs from flat space, we can find schemes where $\kappa =0$ so that these terms do not appear.  

\section{Soft de Sitter Effective Theory}\label{sec:sdset}

From working with interacting scalars in a fixed de Sitter background, we have learned two essential features of EFT in de Sitter.  The first is that cosmological correlators are a reflection of physics at the energy-scale $H$.  This means that flat space power counting is a reasonably good guide for estimating the amplitude of cosmological correlators when we think of $H$ as the only energy scale in the problem.  The second, and less obvious lesson, is that there is still the potential for large corrections coming from lower energies / longer wavelengths, namely where $k/a \ll H$ or $k\tau \ll 1$.  Away from CFTs, our flat space intuition does not tell us how to anticipate results in this regime as it is where the curvature of spacetime is dominant. 

From an EFT point of view, we would like to understand any large long distance effect as being a consequence of power counting, in the usual RG sense.  If the superhorizon theory is far from the behavior of a free theory, then we should be able to identify some relevant or marginal operators which are responsible.  Concretely, our goal is to make the expansion and power counting in $k/(aH) \ll 1$ explicit where relevant, marginal and irrelevant interactions are characterized by scaling with $(k/[aH])^n$ with $n< 0$, $n=0$ and $n>0$ respectively.  To accomplish this goal, we will write an EFT, the Soft de Sitter Effective Theory~\cite{Cohen:2020php} (SdSET), where $[a(t) H] = \Lambda(t)$ is a time-dependent UV cutoff.  We will integrate out modes with hard comoving momenta, $p > [aH]$.

The central observation of SdSET that we will need is that $\phi$ does not behave like a single scaling operator in the long-wavelength limit, $ k \tau \to 0$. If we simply set $k=0$ and solve the equations of motion associated with the massive scalar in Equation~(\ref{eq:massive_scalar}), one finds two solutions 
\beq 
\phi(\k=0, t) = \kappa_\alpha a(t)^{-\alpha}+ \kappa_\beta a(t)^{-\beta}  \ ,
\eeq 
where $\kappa_{\alpha,\beta}$ are constants, $\alpha = \frac{3}{2} - \nu$, and $\beta = \frac{3}{2} +\nu$ so that $\alpha +\beta =3$. Our goal is to define the theory by the scaling with $k/[a(t) H]$ but we see that $\phi$ does not behave as a single scaling operator in the $k \to 0$ limit.  Following the usual EFT playbook, we want to organized our results in terms of explicit powers of the UV cutoff, $\Lambda = a(t) H$ and operators that scale like powers of $k$.  Starting from the original scalar, one may then guess that we want to split the operator $\phi$ into $\vp$ and $\vm$ using the ansatz
\beq\label{eq:ansatz}
\phi  = H \left( \vp(\x, t) [a(t) H]^{-\alpha} + \vm(\x,t)  [a(t) H]^{-\beta} \right) \ .
\eeq
We have factored out the $k=0$ solutions to the equations of motion so that powers of $[aH]$ will appear explicitly in the action after this change of variables (as we will see). In addition, the overall power of $H$ is needed to match the units of $\phi$ as defined by the short distance theory.  With this choice, $\vp$ and $\vm$ have the same scaling and engineering dimensions so that dimensionally $[\vp(\x,t)]= [x]^{-\alpha}$ and $[\vm(\x,t)] = [x]^{-\beta}$. This decomposition matches our expectations from an EFT expansion, as it will obviously break down if we try to extend it beyond the cutoff, $k> aH$, which coincides with subhorizon physics. 

Starting with a free scalar field, we may plug this ansatz into the action to determine the action for $\vp$ and $\vm$:
\beq
S = \int d\t d^3 x \left( -\nu  \dvp \vm +\nu \vp \dvm  + \frac{1}{[aH]^2} \partial_i \vp \partial^i \vm + \ldots \right) \ ,
\eeq
where $\t \equiv H t$.  At order leading-order in derivatives, the resulting equations of motion are simply $\dvpm = 0$, which reflects the fact that our ansatz factored our the solutions to the equations of motion with $k=0$.  Terms with more than one time-derivative, such as ${\dot \varphi}_{\pm}^2$, are removed using the leading equations of motion in order to maintain the correct number of degrees of freedom in the EFT~\cite{Weinberg:2008hq}.  Spatial derivatives always come suppressed by powers of $aH$ and vanish in the limit $aH \to \infty$.

Unlike many EFTs in flat space, the coefficients of the low energy action alone does not specify all the information we need regarding the microscopic (UV) theory.  The action alone tells us that $\vp$ commutes with itself and that the canonical commutator is $[\vp(\x),\vm(\x')] =-i \delta(\x-\x')/(2\nu)$. To match the correlators of the UV theory, we must therefore specify initial conditions for $\vp$ in the form of {\it classical stochastic} variables.  Concretely, the two point function of $\vp$ that follows from Equation~(\ref{eq:mode}) implies that 
\beq
\langle \vp(\k) \vp(\k') \rangle = \frac{C^2_\alpha}{k^3} k^{2\alpha} (2\pi)^3 \delta(\k+\k') \ ,
\eeq
where $C_\alpha=2^{1-\alpha} \frac{\Gamma\left(\frac{3}{2}-\alpha\right)}{\sqrt{\pi}}$, which we defined such that $C_\alpha = 1$ at $\alpha =0$. The overall power of $k^{-3}$ follows from the fourier transform, while the power of $k^{2 \alpha}$ shows that $\vp(\x)$ behaves as a scaling operator of dimension $\alpha$.  In fact, under the $SO(4,1)$ isometries of de Sitter space, $\vpm$ transform as primary operators of dimension $\alpha$ and $\beta$ respectively:
\begin{align}
\varphi_\pm(\x,\t) &\to\left[1-2\Delta_\pm\s x_i b^i + \big(x^{2}-[a H]^{-2}\big) b_i \partial^{i}-2 b_i x^{i} \vec{x} \cdot \vec{\partial} + 2 b_i x^i \partial_{\t} \right]\s \varphi_\pm(\x,\t) \ ,
\label{eq:sc_iso}
\end{align}
where $\Delta_+ \equiv \alpha$ and $\Delta_- \equiv \beta$. The observation that the solutions to the equations of motion behave like primary operators is well-known both in dS and AdS and is central to the (A)dS/CFT holographic dictionary~\cite{Witten:1998qj}. The new development in SdSET is that the bulk action is expressed directly in terms of these operators.

Now, suppose we add interactions to the UV theory in the form of a potential, 
\beq
V(\phi) = \sum_{n>3} \frac{\lambda_n}{n!} \phi^n \ .
\eeq
If we simply apply our ansatz, Equation~(\ref{eq:ansatz}), then the potential expressed in these variables becomes
\beq
V(\phi) = H^4 \sum_{n>3} \sum_{m=0}^{n} [aH]^{-(n-m) \alpha -m \beta} \frac{c_{n-m,m}}{(n-m)! m!}  \vp^{n-m} \vm^m (\x) 
\eeq
where
\beq\label{eq:ltoc}
c_{n-m,m} = \lambda_n H^{n-4} 
\eeq
so that $c_{n-m,m}$ is dimensionless. Notice that if we define $\t = H t$ and $V(\phi)= H^4  V(\vp,\vm)$ then the action, $S \supset \int d\t d^3 x [aH]^3 V(\vp, \vm)$ implements the explicit power counting in $[aH]$ where $\vp$ and $\vm$ carry dimensions $\alpha$ and $\beta$, $\x$ is dimension $-1$, and $\t$ is dimension 0.  We can express this more transparently by expanding in the dimension of the operator, $\Delta_{n,m} = (n-m) \alpha + m\beta$, so that our potential becomes
\bea
S_{n.m} &=& \int d\t d^3 x [aH]^{3-(n-m)\alpha- m \beta} \frac{c_{n-m,m}}{(n-m)! m!}  \vp^{n-m} \vm^m (\x) \nonumber\\
&=& \int d\t d^3 x \frac{1}{\Lambda^{3-\Delta_{n,m}}} \frac{c_{n-m,m}}{(n-m)! m!}  \vp^{n-m} \vm^m (\x) 
\eea
where $\Lambda(t) = a(t)H $ is the (time-dependent) UV cutoff, as desired. 

\begin{figure}[t]
	\begin{center}
		\includegraphics[width=3.5in]{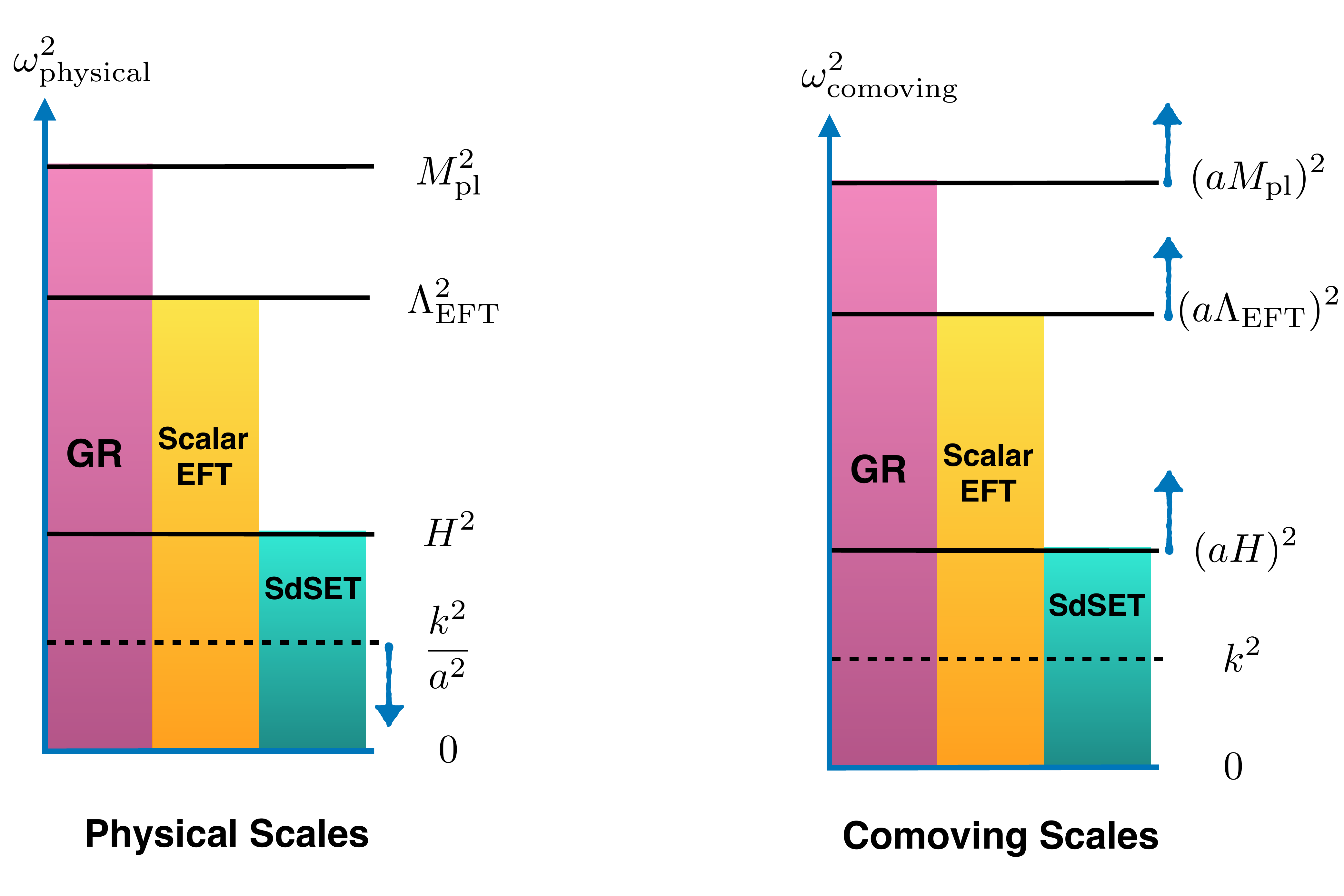}
	\end{center}
	\vskip -.5cm
	\caption{The hierarchy of scales that defines the soft de Sitter effective theory in terms of physical angular frequencies ({\it Left}) and co-moving angular frequencies ({\it Right}). The blue arrows indicate which scales evolve under time-evolution in each description.}
	\label{fig:SdSET_scales}
\end{figure}

So far we have described SdSET from the top down by expressing the short distance field $\phi$ in terms of $\vpm$.  However, like any good EFT, we can derive all the same result from the bottom up using only the degrees of freedom below UV cutoff $\Lambda(t)$, $\vpm$, the symmetries, and the power counting rules.  The special relations between $c_{n-m, m}$ and $c_{n-m',m'}$ that arise from matching to the UV potential, Equation~(\ref{eq:ltoc}), occur in the EFT due to an additional symmetry, reparameterization invariance (RPI), 
\begin{align}
\varphi_{+} &\s\s\to\s\s \varphi_{+} +\epsilon\s [a\s H]^{\alpha-\beta}\s \varphi_{-}  \label{eq:RPITrans} \ ,  \\[5pt]
 \varphi_{-} &\s\s\to\s\s (1-\epsilon)\s\varphi_{-} \ .
\end{align}
This is a symmetry of the low energy theory that keeps the original field, $\phi$, fixed. In addition to symmetries, the rules of EFT also permit us to change variables to remove (redundant) terms in the action.  Interestingly, in SdSET, there are a number of changes of variables that are not field redefinitions of the UV theory. Concretely, we can redefine $\vp$ and $\vm$ independently to remove a number of couplings from the action, yet in the original description one cannot express this as a field redefinition of $\phi$ itself. This additional flexibility is what allows SdSET to simplify and clarify a number of features of physics in dS. 

Naively, power counting suggests that $\vp^n$ is a relevant operator when $n \alpha < 3$, which would imply that the corrections to the scalar fields dynamics grow like powers of $[aH]$. If true, scalar fields with a wide range of masses would be strongly coupled on super-horizon distances in dS.  However, from explicit diagrammatic arguments, it has long been known that perturbative corrections grow at most logarithmically in $aH$~\cite{Weinberg:2005vy,Weinberg:2006ac}. We can understand the absence of power-law growth in the UV description by writing the in-in correlator in the commutator form, Equation~(\ref{eq:inin}) with $\epsilon=0$, 
\begin{align}
\big\langle {\cal O}(\t)\big\rangle&= \sum_{N=0}^{\infty} i^{N} \int_{-\infty}^{\t} \dd \t_{N} \int_{-\infty}^{\t_{N}} \dd \t_{N-1} \cdots \int_{-\infty}^{\t_{2}} \dd \t_{1} \nonumber \\[4pt] 
&\hspace{20pt} \times\Big\langle\big[H_{{\rm int}}\big(\t_{1}\big)\s,\s\big[H_{{\rm int}}\big(\t_{2}\big)\s,\s \cdots\big[H_{{\rm int}}\big(\t_{N}\big)\s,\s {\cal O}^{{\rm int}}\big(\t\big)\big] \cdots\big]\big]\Big\rangle \label{eq:commutator} \ .
\end{align}
Because every insertion of the interaction is associated with a commutator, $[\dot \phi, \phi] \propto a^{-3}$, the overall power of $a^3$ in the measure of integration is always cancelled so that no positive power of $a$ appears alone. This argument is straightforward at tree-level, but is technically complicated for general loop-diagrams.

In the EFT description, the absence of power-law growth to all-loop order arises simply because $\vp^n$ is a redundant operator: it can be removed by a field definition
\beq
\vm \rightarrow \vm+\frac{c_{n, 0}}{9(n-1) !}[a H]^{3-(n-1) \alpha} \vp^{n-1} \ .
\eeq
This is precisely the kind of field redefinition that exists only the EFT and not the the full theory, thus explaining why a complicated result of the full theory can be made trivial in the EFT. As stated above, this does not replace our need to perform the full UV calculations, but is one of several all-orders results that is translated into a simple power counting argument in EFT.

After removing redundant operators, the leading interaction by power counting is $c_{n-1,1}$ which is associated with a dimension $\Delta_{n-1,1}= 3 + (n-2) \alpha$ operator.  This operator is irrelevant ($\Delta_{n-1,1}>3$) for all $\alpha > 0$ and therefore we see that a typical (massive) scalar in dS is a free field in the IR. Similarly, derivative interactions have dimensions $\Delta_{\partial} \geq 5$  for all $\alpha \geq 0$ and are always irrelevant. This leaves the case of massless scalar field, $\alpha=0$, as the unique situation where there could be non-trivial long-distance dynamics on super-horizon scales.

The precise field redefinition needed to remove redundant operators is important for matching, as it modifies the effective potential of the EFT relative to the potential of the UV theory.  Specifically, if we start from $V(\phi) = \lambda \phi^4/4!$ and $\alpha = 0$, then repeated application of this field redefinition to remove all the $c_{n,0}$ terms generates an effective potential in the EFT~\cite{Cohen:2021fzf},
\beq
V_{\rm eff}(\varphi_+,\varphi_-) \supset \frac{\lambda}{3!} \varphi_- \left( \varphi_+^3 + \frac{\lambda}{18}\varphi_+^5 + \frac{\lambda^2}{162} \varphi_+^7 +\,...\, \right) \, . \label{eq:Veff}
\eeq
These higher-order terms are precisely those necessary to match the behavior of the UV theory. The first few terms in this effective potential also reproduce perturbative calculations of the effective potential of light fields calculated using wavefunction~\cite{Gorbenko:2019rza} or static patch techniques~\cite{Mirbabayi:2020vyt}.

In addition to the interactions in the EFT itself, interactions in the UV theory modify the initial conditions of the EFT, in the form of time-independent corrections\footnote{At zeroth order in the EFT expansion, $\dvpm = 0$. As a result, we can always add time-independent initial statistics while remaining consistent the the dynamics of the EFT.} to statistics of $\vp$.    These must be matched to the full theory as they are not predictions of the EFT itself.  For example, if we add a $ (\partial_\mu \phi \partial^\mu \phi)^2$ interaction to the UV theory, as in Equation~(\ref{eq:derivative_int}), then we must match the $\phi$ correlator in Equation~(\ref{eq:4point}) with the $\vp$ initial conditions such that 
\beq
\langle \vp(\k_1,\t) .. \vp(\k_4,\t) \rangle \supset \frac{H^4}{\Lambda^4} \frac{P_6(\k_1,\k_2,\k_3\,\k_4)}{ 16 (k_1 k_2 k_3 k_4)^3 k_t^5} (2\pi)^3 \delta(\sum \k_i)  \ .
\eeq
In this precise sense, the SdSET does not replace the need to perform the UV calculations, as the statistics determined by the UV theory.  However, the key observation is that the initial conditions are time-independent and thus do not influence the time evolution of the fundamental fields, $\vpm(\k,\t)$.

In the previous section, we saw that there can be non-trivial evolution of the superhorizon modes, in the form of anomalous dimensions.  This is important for both massive and massless fields, as we saw they must arise for conformal mass scalars with $m^2 =2 H^2$.  One might wonder how this is consistent with the claim that $\alpha >0$ has only irrelevant interactions. The reason it is consistent is that composite operators require us to integrate over the statistical correlators, including the initial conditions
\beq
\langle \vp^n(\x,\t) \ldots \rangle =\int \left(\prod_{i=1}^n \frac{ d^3 \k_i}{(2\pi)^3} e^{-i \k \cdot \x} \right) \langle \vp(\k_1,\t) .. \vp(\k_n,\t) \ldots \rangle
\eeq
The correlation function on the right-hand side includes the contributions from the initial conditions that are determined by matching. Since the only scale in the problem is $[aH]$, if the $k$-integrals are logarithmically divergent, they will produce factors of $\log k_i /[aH]$.  From the EFT point of view, these are UV divergences and signal the need for (dynamical) RG flow.  This is a standard phenomena in EFT where, in the right variables, IR divergences of the microscopic description are replaced by UV divergences of the EFT~\cite{Cohen:2019wxr}.  The resulting dynamical RG in SdSET gives us an anomalous dimension for $\vp^n$, like the one we found in Equation~(\ref{eq:phi2phi2fullcorrelator}).

Note that the anomalous dimensions are only generated in this way for the composite operators.  In contrast, the dimensions of $\vpm$ are not altered in the EFT, except through interactions in the EFT itself.  The reason is that the dimensions $\alpha$ and $\beta$ are determined by matching and there is no meaningful notion of an anomalous dimension of $\vpm$.  Instead, if such a dimension is generated in the UV (including by a perturbative shift of the effective mass), it is simply absorbed into $\alpha$ by matching.

\section{Light Scalars}\label{sec:light}

Light scalars with $m \ll H$ have long been known to present a significant challenge in de Sitter space. Because their power spectrum is scale-invariant, Equation~(\ref{eq:power}), momentum integrals typically diverge at $k=0$. In addition, even tree-level interactions can give rise to secular growth that needs to be resummed. The framework known as stochastic inflation~\cite{Vilenkin:1983xq,Starobinsky:1986fx,Aryal:1987vn} has long been known to be free of these problems and it has been suggested that it is responsible for resolving these IR issues~\cite{Enqvist:2008kt,Podolsky:2008qq,Seery:2009hs,Seery:2010kh}.  Stochastic inflation translates the freeze-out of each mode of a fundamental scalar field $\phi$ at horizon crossing to a step in a random walk describing the local value of the field. This intuition gives rise to a Fokker-Planck equation for the classical probability distribution for $\phi$, $P(\phi,t)$, in terms of $H$ and the scalar's potential $V(\phi$), 
\beq
\frac{\partial}{\partial t} P(\phi,t)= \frac{H^3}{8 \pi^2} \frac{\partial^2}{\partial \phi^2} P(\phi, t)  + \frac{1}{3 H}  \frac{\partial}{\partial \phi} \big [V'(\phi) P(\phi,t ) \big] \, ,
\label{eq:FPeqStocInf} \ .
\eeq
The first term represents of (Gaussian) quantum noise at horizon crossing while the second term is that classical drift. This equation has been derived from a number of perspectives and has been extended to multi-field models to include non-trivial field-space geometry~\cite{Pinol:2020cdp}. Yet, we have seen that the effective potential $V_{\rm eff}(\vp,\vm)$, Equation~(\ref{eq:Veff}), does not match $V(\phi)$ beyond leading order in $\lambda$ and therefore it is unclear if or how such corrections should be included in the stochastic framework. We would like to understand from EFT power counting how stochastic inflation arises, its regime of applicability, and how to calculate corrections~\cite{Finelli:2008zg,Vennin:2015hra,Markkanen:2019kpv}.

From the action of SdSET, we can observe that theories with massless scalars ($\alpha \to 0$) are the unique situation where the super-horizon theory can have nontrivial IR dynamics. By power counting, potential interactions are marginal when $\alpha \to 0$ and thus do not decouple at late times.  In addition, the massless limit is important because all the composite operators of the form $\vp^n$ have the same dimension in the limit $\alpha \to 0$, $\Delta_{n,0} = n \alpha \to 0$ for all $n$. As operators with the same dimension can mix under RG, this infinite tower of operators can mix in the massless limit, introducing a highly non-trivial RG flow.

The non-trivial mixing of scalar operators is already present the free theory, due the scalar invariance of the power spectrum in Equation~(\ref{eq:power}). We will regulate this divergence by analytic continuation in $\alpha$ and then taking $\alpha \to 0$. From the Wick contraction of two fields, $\alpha \to 0$ gives
\bea
\langle \vp^2(\x) \rangle  = \int \frac{d^3 k}{(2\pi)^3}  \frac{C^2_\alpha}{k^{3-2\alpha}} &\to&  [aH]^{2\alpha} \int \frac{d^3 k}{(2\pi)^3}  \frac{C_\alpha}{(k^2+m^2)^{3/2-\alpha}} \\
&=&\frac{1}{2\alpha} - \log(k/[aH]) +\ldots
\eea
The original integral is both UV and IR divergent, but we regulate the IR, introducing an artificial $m^2$ to isolate the UV divergence.  Including the RG flow from the UV to the IR softens the IR behavior and allows us to take $m^2 \to 0$. After implementing the RG, we will no longer need for the IR regulator as RG resolves in the long-distance behavior.

Combining the divergences from this Wick contraction and the classical evolution of the massless fields, one finds the Callan-Symmanzik equation for this mixing~\cite{Baumgart:2019clc,Baumgart:2020oby}:
\beq
\frac{\partial}{\partial\t}  \big\langle \vp^n(\x) \,...\, \big\rangle =\frac{n(n-1)}{8\s \pi^2} \s \big\langle \vp^{n-2}(\x) \,...\, \big\rangle - \frac{n}{3} \sum_{m>1}\frac{c_{m,1}}{m!}  \big\langle \vp^{n-1}(\x)\s \vp^m(\x) \,...\, \big\rangle  \, . \label{eq:mixing_leading}
\eeq
Note that neither derivatives of $\vp$, nor powers of $\vm$ appear in this equation because they are at least dimension two and three respectively. Given that the non-zero commutator in SdSET is $[\vp,\vm]$, quantum mechanical effects are formally irrelevant to do not impact superhorizon physics (whether or not there is decoherence).  

Now suppose we consider the case of a massless $\phi$ with potential $V(\phi) = \lambda \phi^4/4!$ as the UV theory. SdSET must inherit the $\phi \to -\phi$ symmetry, but we would otherwise expect all possible mixings allowed by this symmetry to arise.  These corrections can also be computed directly in the UV theory, but again we use SdSET with $\vp$ to make the power counting obvious.  Specifically, we know the mixing of operators of the same dimension is not limited in any obvious way and we should expect a sum of the form
\begin{align}
    \frac{\partial}{\partial \t}  \big\langle \vp^n \big\rangle =&- \frac{n}{3} \sum_{m>1}^{\text{odd}}\frac{c_{m,1}}{m!}  \big\langle \vp^{n+m-1}\big\rangle 
+  \binom{n}{2} \sum_{m=0}^{\infty}b_m \big\langle \vp^{n+2m-2} \big\rangle \notag\\[5pt]
&-\binom{n}{3} \sum_{m=0}^{\infty}d_m \big\langle \vp^{n+2m-2} \big\rangle
+\binom{n}{4} \sum_{m=0}^{\infty}e_m \big\langle \vp^{n+2m-4} \big\rangle  + ...\ . 
\end{align}
Interestingly, the information encoded by these equations can be written as a single equation for the probability distribution of $\vp$, 
\begin{align}\label{eq:stocastic_expand}
    \frac{\partial}{\partial \t} P(\vp,\t) &=\frac{1}{3}  \frac{\partial}{\partial\vp} \left [\partial_{\vm}V_{\rm eff} (\vp,\vm)|_{\varphi_-=0} P(\vp,\t ) \right] \notag\\[5pt] 
    &+ \frac{\partial^2}{\partial \vp^2}  \left[ \sum_{m=0}^\infty \frac{ b_m}{2!} \vp^{2m} P(\vp, \t) \right]+  \frac{\partial^3}{\partial \vp^3}  \left(\vp \sum_{m=0}^\infty \frac{d_m}{3!} \vp^{2m} P(\vp, \t) \right) \notag\\[5pt] 
& +  \frac{\partial^4}{\partial \vp^4}  \left( \sum_{m=0}^\infty \frac{ e_m}{4!} \vp^{2m} P(\vp, \t)  \right) + \ldots \ .
\end{align}
This infinite series of terms is typical of a general Markovian process where we can write the time evolution as
\beq\label{eq:markov}
  \frac{\partial}{\partial \t} P(\vp,\t) = \int d\vp' \left( W(\vp|\vp') P(\vp',\t) - W(\vp'|\vp) P(\vp,\t)  \right) 
\eeq
where $W(\vp|\vp')$ is the transition amplitude for a jump from $\vp'$ to $\vp$.  The derivative expansion in Equation~(\ref{eq:stocastic_expand}) is the we result of expanding in $\vp-\vp'$, which is known as the Kramers-Moyal expansion. The coefficients $b_m$ characterized the variance of the Gaussian noise as a function of $\vp$, while the higher order terms like $d_m$ and $e_m$ are the non-Gaussian moments of the transition amplitudes.  

These types of corrections to the stochastic framework can be derived in many ways~\cite{Burgess:2014eoa,Gorbenko:2019rza,Mirbabayi:2019qtx,Mirbabayi:2020vyt}, all of them arising from integrating out modes with momenta $p > aH$. Several of these approaches directly integrate-out our $\vm$ by focusing only on the nearly constant mode~\cite{Burgess:2014eoa,Mirbabayi:2019qtx,Mirbabayi:2020vyt}.  We cannot write a first order action for $\vp$ without $\vm$ but after integrating out $\vm$ one can still arrive at an effective equation of motion like Equation~(\ref{eq:stocastic_expand}) using the language of open EFT~\cite{Burgess:2015ajz}.  From the perspective of SdSET, we would expect this procedure only works when $\alpha \approx 0$ so that $\vm$ and $\vp$ don't mix. This description has the advantage of making the connection to the thermal behavior of the static patch more explicit~\cite{Anninos:2011af,Mirbabayi:2019qtx,Mirbabayi:2020vyt}.  Alternatively, one can work with the wavefunction of the universe as an intermediate step~\cite{Gorbenko:2019rza}, which is advantageous for isolating the origin of the logarithmic terms~\cite{Anninos:2014lwa}. Similar results can be derived directly from the in-in expression by diagrammatic arguments~\cite{Baumgart:2019clc,Baumgart:2020oby}. In a certain sense, all these approaches behave as EFTs in that they remove the subhorizon physics. 

The concrete advantage of SdSET is that it reduces the problem to calculating the matrix of anomalous dimensions from scaleless loop integrals. In this precise sense, calculating the coefficients of the Kramers-Moyal expansion is essentially identical to finding the dimensions of operators at the Wilson-Fisher fixed point in $d=4-\epsilon$ dimensions. The first higher derivative term, e.g.~$d_0$ in $\lambda\phi^4$, has only be calculated using this method, which at the appropriate order in $\lambda$ gives the corrected equation~\cite{Cohen:2021fzf}
\begin{align}
\hspace{-10pt}\frac{\partial}{\partial \t} P(\bvp,\t)
    &= \frac{1}{3}  \frac{\partial}{\partial\bvp} \big [V'_{\rm eff}(\bvp) P(\bvp,\t ) \big] 
    +\frac{1}{8\pi^2}\frac{\partial^2}{\partial \bvp^2} P(\bvp, \t) \notag \\
   &+ \frac{\lambda_{\rm eff}}{1152\pi^2}  \frac{\partial^3}{\partial \bvp^3}  \big(\bvp   P(b\bvp, \t) \big)
    \label{eq:FPatNNLO}\\[10pt]
   \hspace{55pt} V'_{\rm eff} &= \frac{\lambda_{\rm eff}}{3!} \bigg( \bvp^3 + \frac{\lambda_{\rm eff}}{18}\bvp^5 + \frac{\lambda_{\rm eff}^2}{162} \bvp^7 +\,...\, \bigg)\,.
    \label{eq:VpEff}
\end{align}
where $\lambda \to \lambda_{\rm eff}$ and $\vp \to \bvp$  were redefined to remove the $b_1$ and $b_2$ terms.

Despite the advantages of the SdSET approach, it remains tied to the UV calculation via the matching of initial conditions. Other approaches may be able to circumvent this technical requirement while maintaining power counting. For example, Mellin-space makes loop-integrals scaleless and can simplify some of the technical complications of regulating the UV description~\cite{Premkumar:2021mlz}. In addition, objects like the wavefunction of the universe~\cite{Heemskerk:2010hk,Harlow:2011ke} might be more natural starting points for non-perturbative dynamical RG~\cite{Cespedes:2020xqq,Goodhew:2020hob}, in analogy with the exact RG results from the path integral. SdSET is a continuum EFT, in the sense of~\cite{Georgi:1993mps}, which has technical advantages for concrete calculations, but lacks the conceptual and non-perturbative advantages of Wilsonian EFT. 

Given the expansion in derivatives with respect to $\vp$ in our effective Fokker-Planck equation, Equation~(\ref{eq:FPatNNLO}), it is reasonable to wonder what parameter controls the size of the higher derivative terms. To gain intuition, let us start by truncating the equations at two derivatives and linear order in $\lambda_{\rm eff}$, namely the Fokker-Planck equation
\beq\label{eq:FP}
\frac{\partial}{\partial \t} P(\bvp,\t) = \frac{1}{3}  \frac{\partial}{\partial\bvp} \big [ \frac{\lambda_{\rm eff}}{3!} \vp^3 P(\bvp,\t ) \big] +\frac{1}{8\pi^2}\frac{\partial^2}{\partial \bvp^2} P(\bvp, \t) \ .
\eeq
We can calculate the equilibrium probability distribution~\cite{Starobinsky:1994bd}, $\dot P^{\rm FP}_{\rm eq}(\vp) = 0$ by direct integration to find
\beq\label{eq:FP_sol}
P^{\rm FP}_{\rm eq}(\vp) = \exp\left(- \frac{\pi^2}{9} \lambda_{\rm eff} \vp^4 \right) \ .
\eeq
Since the probability is of order one up to $|\vp| \sim \lambda_{\rm eff}^{-1/4}$, we see that the counting in $\lambda_{\rm eff} \ll 1$ is modified by large values of $\vp$ in equilibrium. Using the power counting $\vp \sim \lambda_{\rm eff}^{-1/4}$, the leading order behavior (LO) of Equation~(\ref{eq:FPatNNLO}), ${\cal O}(\lambda_{\rm eff}^{1/2})$, is given by standard Stochastic Inflation, as in Equation~(\ref{eq:FP}).  At next-to-leading order (NLO), ${\cal O}(\lambda_{\rm eff})$, only the $\vp^5$ term in $V'_{\rm eff}(\vp)$ contributes, with the remaining terms being next-to-next-to-leading order (NNLO), ${\cal O}(\lambda_{\rm eff}^{3/2})$. We can write the equilibrium probability distribution as a similar expansion in $\lambda_{\rm eff}$,
$P_{\rm eq} = C P_{\rm LO}(\vp) P_{\rm NLO}(\vp) P_{\rm NNLO}(\vp)$, such that the solution to NNLO is~\cite{Cohen:2021fzf}
\begin{align}
P_{\rm LO} &= \exp\left(- \frac{\pi^2}{9} \lambda_{\rm eff} \vp^4 \right)\\[4pt] 
P_{\rm NLO} &= \exp\left(- \frac{\pi^2}{243} \lambda_{\rm eff}^2 \vp^6 \right) \\[4pt]
P_{\rm NNLO} &=  \exp\left( \frac{5}{10368} \lambda_{\rm eff}^2 \vp^4 - \frac{17\pi^2}{46656} \lambda_{\rm eff}^3 \vp^8 \right)\,.
\end{align}
The same power counting can be applied to the relaxation eigenvalues,
\beq
\frac{d}{dt} P_i(\varphi)  = - \Lambda_i P_i(\varphi) \ ,
\eeq
which have been similarly calculated to NLO~\cite{Gorbenko:2019rza,Mirbabayi:2020vyt} and NNLO~\cite{Cohen:2021fzf}.

The key take-away is that, from a number of distinct perspectives~\cite{Burgess:2014eoa,Anninos:2014lwa,Gorbenko:2019rza,Mirbabayi:2019qtx,Baumgart:2019clc,Baumgart:2020oby,Cohen:2020php,Mirbabayi:2020vyt,Cohen:2021fzf}, the IR divergences and secular growth of the massless scalar fields in FRW slicing of dS are now understood. EFT trades the bad IR behavior for UV divergences in the usual sense.  The UV divergences give rise to an RG (or equivalently, stochastic inflation) and show that the theory flows to a non-trivial fixed point (equilibrium distribution) where $\varphi \sim \lambda^{-1/4}$. This has long been conjectured as the resolution~\cite{Enqvist:2008kt,Podolsky:2008qq,Seery:2009hs,Seery:2010kh}: stochastic inflation is free of the IR problems of in-in perturbation theory and thus would solve the IR issues if it was equivalent to QFT. What recent works have demonstrated is how to derive this result from QFT and how to calculate corrections to these results to any order in $\lambda$. More significantly, as we now understand the origin of these challenges in terms of power counting, it is straightforward to generalize these results to any interacting QFT in de Sitter.  

\section{Dynamical Gravity}\label{sec:metric}

Dynamical gravity presents an interesting paradox for pure de Sitter and inflationary backgrounds.  It is well-known that backreaction of scalar fluctuations changes the nature of the spacetime; in the most extreme case, this gives rise to eternal inflation~\cite{Steinhardt:1982kg,Vilenkin:1983xq,Linde:1986fd,Guth:2007ng} where even our qualitative understanding in limited.  Yet, in perturbation theory, both the scalar and tensor modes are conserved outside the horizon and are much better behaved than most QFTs on a fixed dS background. 

Because the metric fluctuates at the future boundary of de Sitter, one might worry that the lack of a physical observable makes the discussion of correlators of metric fluctuations meaningless~\cite{Witten:2001kn,Bousso:2004tv}. Working instead in an inflationary background helps clarify the role of observables, as the background scalar field defines a reference frame in which to compute well-defined correlators that are relevant to observations in our universe. We can work in a gauge where this background field is homogenous, $\phi(\x,t) = \phi(t)$, and the fluctuations are encoded in the metric,
\beq
d s^2=-N^2 d t^2+a^2(t) e^{2 \zeta(\vec{x}, t)}\left(e^{\gamma(\vec{x}, t)}\right)_{i j}\left(d x^i+N^i d t\right)\left(d x^j+N^j d t\right) \ ,
\eeq
where $\zeta$ is the scalar (adiabatic) mode and $\gamma_{ij}$ are the tensor modes with $\gamma_i^i=0$ and $\partial_i \gamma^i{ }_j=0$.

Inflation as a framework is far more general than just the rolling of a scalar field. The inflationary epoch itself characterize by an EFT, the EFT of Inflation~\cite{Cheung:2007st}, when three conditions are satisfied
\begin{itemize}
\item The expansion was nearly exponential so that the Hubble parameter, $H(t)  = \dot a / a$, was nearly constant, $|\dot H| \ll H^2$. This condition allows for the creation of scale-invariant long-wavelength fluctuations.
\item There was a physical clock that defined a preferred time slicing.  This condition is necessary to allow inflation to end, starting the hot thermal evolution and allowing the long-wavelength fluctuations to seed structure.
\item The preferred slicing manifests itself in the flat space limit\footnote{This condition is needed to distinguish the EFT of Inflation from models like Solid Inflation~\cite{Endlich:2012pz}, where the cosmological background is the source of the time dependence.} as an operator that breaks time-translation, $\langle {\cal O}\rangle \propto t$. Inside the horizon, the fluctuations of the metric are characterized by the Goldstone boson, $\pi$, of the time-translation breaking, through the relation $\zeta = - H \pi +{\cal O}(\pi^2)$.
\end{itemize}
A wide variety of models of inflation can be made satisfying these criteria, some of which have already been excluded by precision observations of the CMB~\cite{Planck:2018jri,Planck:2019kim}. The EFT of Inflation is particularly useful for calculating the statistics of the fluctuations in $\zeta$ as they are simply determined from the action for $\pi$ on a fixed background (to the required accuracy for current observations). In contrast, for determining the superhorizon behavior of the metric in an inflationary background, $\zeta$ itself is the more important variable. For our purposes, we can pick the gauge $\pi =0$ so that we are working directly with the metric fluctuations, but at the cost of limiting the utility of the EFT of Inflation.

The principle origin of the simplicity of metric fluctuations during inflation is that, after gauge fixing the small diffeormorphisms, the metric remains invariant under an infinite set of large diffeomorphisms that act as symmetries of the long distance theory~\cite{Hinterbichler:2013dpa}.  The simplest such symmetry is a constant rescaling of coordinates, $\x \to e^{-\lambda} \x$, that shifts of the adiabatic mode by a constant $\zeta \to \zeta -\lambda$. This symmetry implies that a physical long-wavelength adiabatic mode is locally indistinguishable from a change of coordinates, up to gradients that vanish as $(k/(aH))^2$. This observation is also essential to our conceptual understanding of inflationary fluctuations, including the separate universes approach to calculating cosmological coorelators~\cite{Salopek:1990jq,Wands:2000dp}. In addition,  this symmetry requires that a constant mode is necessarily a solution to the equations of motion~\cite{Weinberg:2003sw}, which is central to the conservation of $\zeta$ around any FRW background~\cite{Salopek:1990jq}.

More recently, this set of all such symmetries acting on physical fluctuations has been classified. For the adiabatic mode, the key symmetries that act on $\zeta$ are~\cite{Hinterbichler:2012nm}  
\begin{align}
D_{\mathrm{NL}}: \delta \zeta&=-1-\vec{x} \cdot \vec{\partial}_{\vec{x}} \zeta \label{eq:DNL} \\[5pt]
K_{\mathrm{NL}}^{i}: \delta \zeta&=-2 x^{i}-2 x^{i}\Big(\vec{x} \cdot \vec{\partial}_{\vec{x}} \zeta\Big)+x^{2} \partial^{i} \zeta \label{eq:KNL}\ ,
\end{align}
which form a group of nonlinearly realized conformal transformations, $SO(4,1)$, such that $\zeta$ acts like a dilaton. The symmetries acting on the tensors are more complicated, but can be written in terms of a large diffeomorphism, $\x\to \x+\vec \xi$, and
\beq
\delta \gamma_{i j}=\partial_{i}\s \xi_{j}+\partial_{j}\s \xi_{i} \ ,
\eeq
where
\beq
\xi^{(M)}_{i}=M_{i \ell_{1}}\s x^{\ell_{1}}+\frac{1}{2}\s M_{i \ell_{1} \ell_{2}}\s x^{\ell_{1}}\s x^{\ell_{2}}+\frac{1}{3 !}\s M_{i \ell_{1} \ell_{2} \ell_{3}}\s x^{\ell_{1}}\s x^{\ell_{2}}\s x^{\ell_{3}}+\ldots \ ,
\eeq
and $M_{i \ell_1..\ell_n}$ are constant tensors that obey a limited set of constraints~\cite{Hinterbichler:2013dpa}.

Our understanding of the long wavelength behavior has been formalized using this full group of symmetries.  Inside correlators, they manifest themselves as Ward identities that constraint the soft behavior, including Maldacena's single-field consistency condition~\cite{Maldacena:2002vr},
\beq
\lim_{\k_1 \to 0} \langle  \zeta(\k_1)  \zeta(\k_2) \zeta(\k_3)  \rangle \to  -(n_s-1) P(k_1) P(k_3) (2\pi)^3 \delta(\sum_i \k_i) \ .
\eeq
In short, this expression tells us that long modes decouple from short modes, up to an overall coordinate change~\cite{Pajer:2013ana,Dai:2015jaa}.  Acting directly on the operators, symmetries and consistency conditions are also essential in extending the proof of the conservation of $\zeta$ to all-loop order.  One can use the consistency condition inside individual diagrams to restrict the form of loop corrections~\cite{Pimentel:2012tw,Senatore:2012ya}. Alternatively, one can use the symmetries to write an operator equation of $\zeta$~\cite{Assassi:2012et}
\beq
\frac{d}{dt} \zeta(\x,t) = \sum_\Delta {\cal O}_\Delta(\x,t) = \frac{1}{[aH]^2} (c_1 \partial^2 \zeta + c_2 \partial_i \zeta \partial^i \zeta) + \ldots \ ,
\eeq
where the constant $c_{1,2}$, for example, are model dependent. Because $\zeta$ transforms under Equation~(\ref{eq:DNL}) nonlinearly (i.e.~a shift symmetry), any operators ${\cal O}_\Delta$ written in terms of $\zeta$ must contain derivatives and are therefore suppressed by powers of $aH$. This also explains why conservation of $\zeta$ fails in multifield inflation, where  ${\cal O}_\Delta$ can be a non-derivative operator containing a spectator field. 

Unfortunately, in the UV description, showing that the power counting applies to ${\cal O}_\Delta$ in a generic correlation function requires a complicated diagrammatic argument~\cite{Assassi:2012et}. Without the EFT, we cannot simply count powers of $aH$ to determine the behavior of an operator. With the benefit of hindsight, we know corrections that ruin power counting would have to contribute positive powers of $[aH]$ to cancel the explicit negative powers.  Such conributions correspond to power law divergences in the SdSET and can be removed by an appropriate change of operator basis.

By comparison, conservation of $\zeta$ and $\gamma$ in SdSET is a relatively trivial observation about the dimension $\alpha$ associated with $\zeta_+$ or $\gamma_+$, defining $\zeta_+, \gamma_+$ by Equation~(\ref{eq:ansatz}) with $\phi \to \zeta, \gamma$. Concretely, the nonlinearly symmetries must act on the growing modes $\zeta_+$ and $\gamma_+$,  to be consistent with the long wavelength limit, $\zeta \to H \gamma_+$ and $\gamma \to H \gamma_+$; yet this symmetry is only possible when $\alpha = 0$ for both $\gamma_+$ and $\zeta_+$.  In addition, the EFT cannot contain non-derivative interactions and therefore $\zeta$ and $\gamma$ are conserved as operators. This also provides the first all-orders proof of the conservation of $\gamma_{ij}$. Altogether, this implies that in single field inflation
\beq
\lim_{\k \to 0} \langle \dot \zeta(\k) ... \rangle \to 0 \qquad \lim_{\k \to 0} \langle \dot \gamma_{ij}(\k)... \rangle  \to 0 \ ,
\eeq
for any correlation function. Note that this is far more restrictive than the behavior of a massless scalar field in de Sitter space. However, the equations are the result of a nonlinearly realized large diffeomorphism and thus will not be apparent for individual terms in the action of $\zeta$ or if we introduce a regulator that breaks these symmetries. As discussed in Section~\ref{sec:EFT}, there are few good regulators in de Sitter which is why these all loop results are not obvious in many direct calculations in the UV description. It would be interesting to revisit the all-orders conservation in Melin space where many of these challenges are mitigated~\cite{Premkumar:2021mlz}.  

From our discussion of light scalar fields, one is also naturally interested in the time evolution of composite operators. Writing the most general local equation and applying the symmetries in Equations~(\ref{eq:DNL}) and~(\ref{eq:KNL}), one finds that the time evolution of $\zeta^N$ must be governed by~\cite{Cohen:2021jbo}
\beq
\frac{\partial}{\partial\t} \zeta^N(\x,\t) =  \sum^N_{n\geq 2} \gamma_{n} \left(\!\!\begin{array}{c}
N \\
n
\end{array}\!\!\right) \zeta^{N-n}(\x,\t) \ .
\eeq
Note that for $N=1$, the right-hand-side is zero as needed for conservation of the long modes.  Translating this to stochastic inflation gives a generalization of the Fokker-Planck equation, 
\beq
\frac{\partial}{\partial\t} P(\zeta,\t) = \sum_{n\geq 2} (-1)^n \frac{\gamma_n}{n!} \frac{\partial^n}{\partial\zeta^n} P(\zeta, \t) \ .
\label{eq:EvolutionPofZetaGeneral}
\eeq
Note that, in contrast to Equation~(\ref{eq:stocastic_expand}), only derivatives of $\zeta$ appear, as we would expect from the shift symmetry. The coefficients $\gamma_n$ are determined from integrating the connected $n$-point function
\beq
\langle \zeta^n(\x =0)\rangle \propto \gamma_n \log [aH]
\eeq
In this concrete sense, the coefficients $\gamma_{n>2}$ characterize the non-Gaussian noise associated with horizon crossing.

Unlike an interacting scalar with a potential $V(\phi)$, the solution to the Fokker-Planck equation does not lead to an equilibrium solution.  For a free theory, with $\gamma_{n>2}= 0$ and $\gamma_2\equiv \sigma^2 = \Delta_\zeta /(2\pi^2)$, one finds a solution
\beq\label{eq:gauss_sol}
 P_\text{G}(\zeta,\t,\zeta_0) = \frac{1}{\sqrt{2\pi \sigma^2 \t} } e^{- (\zeta-\zeta_0)^2/(2 \sigma^2 \t)} \ .
\eeq
Note that this is nothing other than the solution for the evolution of a Gaussian random walk with variance $\sigma$ and initial conditions $\zeta = \zeta_0$ at $\t=0$, as one would expect from the Fokker-Planck equation. In this sense, it is not surprising that there is no equilibrium solution: each mode crosses the horizon and add a random shift in $\zeta$ which adds coherently because $\zeta(\k)$ freezes outside the horizon.

For an interacting theory, the random walk behavior is generalized to include non-Gaussian noise, as we might have expected from the general Markovian evolution in Equation~(\ref{eq:markov}).  Introducing $\gamma_{n>2} \neq 0$, we can still find a general solution with initial conditions $\zeta = \zeta_0$ at $t=0$,
\beq
P(\zeta;\t,\zeta_0) = \exp\left(\sum_{n>2} (-1)^n \, \frac{\gamma_n \t}{n!} \frac{\partial^n}{\partial \zeta^n} \right) P_\text{G}(\zeta;\t,\zeta_0)  \ ,
\eeq
Notice that around the peak of the distribution, $|\zeta-\zeta_0| \sim \sigma \sqrt{\t}$, we have for $n>2$ and $\t \gg 1$,
\beq
\frac{\gamma_n \t}{n!} \frac{\partial^n}{\partial \zeta^n} P_\text{G}(\zeta;\t,\zeta_0)  = {\cal O}(\gamma_n \t^{1- n/2} \sigma^{-n})  P_\text{G}(\zeta;\t,\zeta_0) \ll  P_\text{G}(\zeta;\t,\zeta_0)
\eeq
We see that after a large number of efolds ($\t \gg 1$) the distribution tends to a Gaussian, as we would expect from the central limit theorem.  As a result, the behavior of the probability distribution of $\zeta$ using stochastic inflation is again improved compared to standard perturbation theory. In this case, rather than a non-trivial fixed point in the form of a equilibrium distribution, the fluctuations of $\zeta$ approach the behavior of a Gaussian random walk\footnote{We can understand the difference between the two cases from the fact that the potential is a relevant deformation of a random walk.}.  Importantly, even though the probability distribution is a Gaussian, it is distinct from the probability distribution of the in-in theory due to the additional by powers of $\t$.

Implicit in this discussion is that we can define the time, $\t$, at which the fluctuations are measured. For inflation itself, when the background classical evolution of the scalar field defines the end of inflation, we can define observables on the $\phi = {\rm constant}$ surface, which characterizes the end of inflation or reheating. However, when the quantum fluctuations of $\phi$, or alternatively $\zeta$, become large enough to overwhelm the classical evolution, the lack of non-perturbative observables in de Sitter space ultimately limits our understanding. It was argued in~\cite{Creminelli:2008es} that there is a phase transition that occurs between these two regimes, where the volume of the reheating surface, $V$, is the order parameter.  When $\langle V \rangle < \infty$, given a finite initial volume, then the reheating surface is well defined and statistics on that surface are well-defined in the semi-classical sense.  As we change the parameters of inflation such that $\langle V \rangle \to \infty$, eternal inflation occurs and the reheating surface is no longer well-defined. The benefit of this approach is that one can define the onset of eternal inflation from the regime that is under control, without requiring that we define observables directly in the eternally inflating regime.

The phase transition to eternal inflation can be defined using the above probability distribution.  In the limit of vanishing slow-roll parameters, the relationship between the evolution of the scalar field driving inflation and the metric fluctuation becomes $\phi = \dot \phi_0 (\t - \zeta )/H$ where $\dot \phi_0$ is a constant.  The probability that reheating occurs at time $\t$, $p_{\rm R}(t)$, is determined the number of points that reach the end point of inflation, $\phi = \phi_c$, at time $\t$ that had previously been inflating, $\phi < \phi_c$, for all $\t' < \t$.  To simplify the discussion, we can set $\phi_c=0$ and therefore the end of inflation corresponds to $\zeta = \t$. Assuming we can neglect all the non-Gaussian in terms in Equation~(\ref{eq:EvolutionPofZetaGeneral}), the probability distribution for $\phi$ or $\zeta$ is given by Equation~(\ref{eq:gauss_sol}) but where we impose absorbing boundary conditions at $\zeta =\t$, as inflation ends when $\zeta$ reaches this point,
\begin{align}
P_\text{G}(\phi[\zeta]<0, \t,\zeta_0) &= \frac{1}{\sqrt{2\pi \sigma^2 \t} } \left[e^{- (\zeta-\zeta_i)^2/(2 \sigma^2 \t)} - e^{-4 \zeta_i / (2\sigma^2)} e^{- (\zeta+\zeta_0)^2/(2 \sigma^2 \t)} \right] \ ,
\end{align}
where again $\zeta_0$ is the initial value of $\zeta$ at $\t =0$.  Following~\cite{Creminelli:2008es}, the reheating probability is
\beq
p_{\rm R}(\t)  = -\frac{\d}{\d \t} \int^{0}_{-\infty} \d\phi\s P_\text{G}(\phi;\t) = -\frac{\d}{\d \t} \int^{\phi_c}_{-\infty} \d\zeta P(\zeta;\t) \propto e^{-\t / (2\sigma^2)} \ ,
\eeq
where we used the Fokker-Planck equation and integrated by parts. The order parameter for eternal inflation is the average reheating volume, 
\beq
\langle V \rangle_\text{G} = L^3 \, \int_{0}^{\infty} \d\t\s e^{3 \t}\s p_\text{R,G}(t)  \simeq  L^3 \, \int_{0}^{\infty} \d \t\s e^{\t (3 - 1/(2\sigma^2))} \ ,
\eeq
where $L^3$ is the size of the initial patch at $\t=0$.   One then defines the phase transition to eternal inflation as the value of $\sigma$ (or $H^2/\dot \phi_0$) where the average volume of the reheating surface diverges, or
\beq
\sigma^2 = \frac{\Delta_\zeta}{2\pi^2}> \frac{1}{6} \ .
\eeq
Although this does not give us any direct insight into the correct physical description of the eternal inflating regime, it gives us a sharp notion of where the boundary lies which controls our observables when the reheating volume is finite.

Naturally, one might wish to revisit this calculation in the presence of non-trivial interactions during inflation. Naively, one might imagine the result is largely unchanged, again due to the central limit theorem. However, the eternal inflation regime is defined by $|\zeta -\zeta_0| \propto \t$, which is not a typical fluctuation. If we repeat our power counting in this regime, we have for $n>2$ and $\t \gg 1$,
\beq
\frac{\gamma_n \t}{n!} \frac{\partial^n}{\partial \zeta^n} P_\text{G}(\zeta;\t,\zeta_0)  = {\cal O}(\gamma_n \t \sigma^{-2n})  P_\text{G}(\zeta;\t,\zeta_0)  \ .
\eeq
This contribution can easily dominate over the Gaussian term, even a model with small fluctuations $\sigma \ll 1$ and is weakly coupling at horizon cross ($\gamma_n \sigma^{-n} \ll 1$).  These models are known to have weakly coupled UV completions~\cite{Baumann:2011su} which suggests the breakdown is not due to strong coupling in the model of inflation but is instead a breakdown of the EFT description of inflation or stochastic inflation when calculating the tail of the probability distribution.  This phenomena is not unique to cosmology but is a reflection of the large deviation principle~\cite{TOUCHETTE20091}, which characterizes random walks that are linear in time (or the number of steps). These large fluctuations are sensitive to the microscopic details of the walk and do not follow the central limit theorem. Improving our understanding of rare fluctuations and the origin of the breakdown of EFT is important for both theoretical questions about cosmology, like the nature of eternal inflation and the meaning of the de Sitter entropy~\cite{Dubovsky:2008rf}, and observational consequences, like the production rate of primordial black holes~\cite{Vennin:2020kng}.

\section{Conclusions}\label{sec:conclusions}

The physics of de Sitter space presents a significant challenge to our understanding of quantum gravity~\cite{Witten:2001kn,Arkani-Hamed:2007ryv,Susskind:2021yvs}. There is no boundary where metric fluctuations decouple in which we can define observables, raising the question of whether there are any non-perturbative observables in a quantum theory of de Sitter space~\cite{Bousso:2004tv,Anninos:2012qw}.  It has been suggested that the more basic technical challenges associated with de Sitter, like IR divergences and secular growth, are tied to these serious non-perturbative questions, even though they are not necessarily related.

Effective field theory gives us a more precise language to discuss the successes and failures of QFT and quantum gravity in de Sitter or inflationary backgrounds. Quantum field theory and perturbative quantum gravity on a fixed de Sitter background is characterized by a single energy scale, $H$, the rate of expansion in the cosmological slicing. The influence of interactions on the statistics of the long-wavelength fluctuations are determined by flat space power counting, where $H$ plays the role of the center of mass enter.  This can also be understood from the perspective of the static patch, where the local physics looks thermal with an effective temperature $T_{\rm dS} = H/(2\pi)$~\cite{Gibbons:1977mu}.  From the EFT point of view, the litany of problems often associated with de Sitter itself are often just the result of lacking good regulators for QFT in cosmological backgrounds. Without such a regulator, short-distance power-law divergences introduce order-one shifts to the couplings and require careful treatment. Recent progress in calculating cosmological correlators using Mellin space~\cite{Sleight:2019mgd,Sleight:2019hfp,Premkumar:2021mlz} or from analytic continuation from AdS~\cite{Meltzer:2021zin,Sleight:2021plv} might circumvent these challenges and make the implementation of basic EFT in dS straightforward.

While better regulators can resolve challenges associated with short distances, it does not fully explain all the divergences of loop corrections at late times and large distances. In the language of EFT, any such large effect should be understandable in terms of power counting, where interactions that are important (negligible) on super-horizon scales are relevant (irrelevant) in the renormalization group sense.  This is not possible in the conventional QFT description and requires a rewriting the theory in terms of scaling operators of the superhorizon description. The soft de Sitter effective theory provides such a description where we can understand phenomena ranging from stochastic inflation to conservation of the adiabatic modes in terms of power counting and symmetries.

The ultimate questions about de Sitter and inflation are expected to take us far from a fixed background with small fluctuations.  Eternal inflation~\cite{Steinhardt:1982kg,Vilenkin:1983xq,Linde:1986fd,Guth:2007ng}, for example, is a poorly characterized cosmological phase where it is difficult to even define observables~\cite{Freivogel:2011eg}.  Part of the power of EFT is that it predicts its own demise, usually in terms of simple power counting. By reshaping our understanding of de Sitter to make power counting manifest, we hope to make the challenges that remain more transparent and familiar. In addition, by isolating the essential long distance degrees of freedom, these problems may appear more trackable and open the door to new approaches and solutions.

\end{document}